\documentclass{aastex}

\begin{document}

\title{\emph{HST} CALSPEC FLUX STANDARDS: SIRIUS (AND VEGA)}
\author{R.~C.\ Bohlin} 
\affil{Space Telescope Science Institute, 3700 San Martin Drive, Baltimore, MD 21218}
\email{bohlin@stsci.edu}
% received 2014Feb10, resubm 2014Mar13

\begin{abstract} The Space Telescope Imaging Spectrograph (STIS) has measured
the flux for Sirius from 0.17--1.01~$\mu$m on the \emph{HST} White Dwarf scale.
Because of the cool debris disk around Vega, Sirius is commonly recommended as
the primary IR flux standard. The measured STIS flux agrees well with
predictions of a special Kurucz model atmosphere, adding confidence to the
modeled IR flux predictions. The IR flux agrees to 2--3\% with respect to the
standard template of Cohen and to 2\% with the MSX absolute flux measurements in
the mid-IR. A weighted average of the independent visible and mid-IR absolute
flux measures implies that the monochromatic flux at 5557.5~\AA\ (5556~\AA\ in
air) for Sirius and Vega, respectively, is $1.35\times10^{-8}$ and
$3.44\times10^{-9}$~erg cm$^{-2}$ s$^{-1}$ \AA$^{-1}$ with formal uncertainties
of 0.5\%. Contrary to previously published conclusions, the Hipparcos photometry
offers no support for the variability of Vega. Pulse pileup severely
affects the Hp photometry for the brightest stars. \end{abstract}  


\section{INTRODUCTION}

Precise stellar flux standards are required for the calibration of the James
Webb Space Telescope (JWST) and for the interpretation of dark energy measures
with the supernova Ia technique. Cohen et al. (1992a) and, more recently,
Engelke et al. (2010, EPK) recommend the use of Sirius as the primary IR
standard, because Vega's rapid rotation and dust ring complicate the
modeling of its IR flux distribution. Thus, Sirius ($\alpha$~CMa, HD~48915,
HR~2491) was observed by \emph{HST/STIS} on 2012 Oct 7 and 2013 Jan 26. The hot
WD companion, Sirius B, is 10 mag fainter at V and contributes $<$1\% of the
system flux, even at 1300~\AA\ (Holberg et al. 1998, Beuermann et al. 2006).

The HST flux system (Bohlin \& Gordon 2014) is based on the flux distribution of
NLTE model atmospheres for the pure hydrogen white dwarfs (WDs) GD153 and GD71
and on a NLTE metal line-blanketed model of Rauch et al. (2013, RWBK) for
G191B2B. The absolute normalization of each model flux is defined by the STIS
net signal in electrons/s from each WD relative to the STIS net signal for Vega
at 5557.5~\AA\ (5556~\AA\ in air), where Megessier (1995) found an absolute flux
of $3.46\times10^{-9}$~erg cm$^{-2}$ s$^{-1}$ \AA$^{-1}\pm0.7$\%. This paper
reconciles the Megessier visible flux with the MSX mid-IR fluxes and derives
$3.44\times10^{-9}$~erg cm$^{-2}$ s$^{-1}$ \AA$^{-1}\pm0.5$\% at 5556~\AA\ for
Vega's monochromatic flux. This 0.6\% change to the HST fluxes also brings the
extrapolated flux for Sirius to within 0.6\% of the average MSX mid-IR absolute
flux measures.

The STIS Sirius observations and their flux calibration are discussed in
Section~2. Section~3 compares the modeled IR spectral energy distribution (SED)
with the MSX absolute flux measurements, while Section~4 discusses Vega, its
dust rings, and the lack of any evidence for variability in the Hipparcos data.

\section{CALIBRATION OF THE SATURATED OBSERVATIONS} %2

STIS observations of Sirius in the three CCD low dispersion modes G230LB, G430L,
and G750L (Hernandez 2012) heavily saturate the full well depth of the CCD
detector. However, the excess charge just bleeds into adjacent pixels along the
columns perpendicular to the dispersion axis and is not lost at Gain~=~4.
Gilliland et al. (1999, GGK) demonstrated that saturated signals on the STIS CCD
are linear in total charge vs. stellar flux, as long as the extraction region on
the image is large enough to include all the charge. In particular, GGK
demonstrated linearity to 0.1\% accuracy using 50$\times$ overexposed images of
a star in  M67 compared with unsaturated exposures of the same star.

Sirius data extraction and calibration proceeded as detailed for
similarly saturated observations of Vega (Bohlin \& Gilliland 2004, BG), except
that taller extraction heights of 206, 182, and 148 pixels are required for
G230LB, G430L, and G750L, respectively, for single 4s exposures with G230LB and
0.3s for G430L and G750L. For these saturated images, the signal level is so
high that any signal loss due to charge transfer efficiency (CTE) effects
(Goudfrooij \& Bohlin 2006) is $<$0.1\%.

\begin{figure}%1 
\centering 
\includegraphics[height=6in]{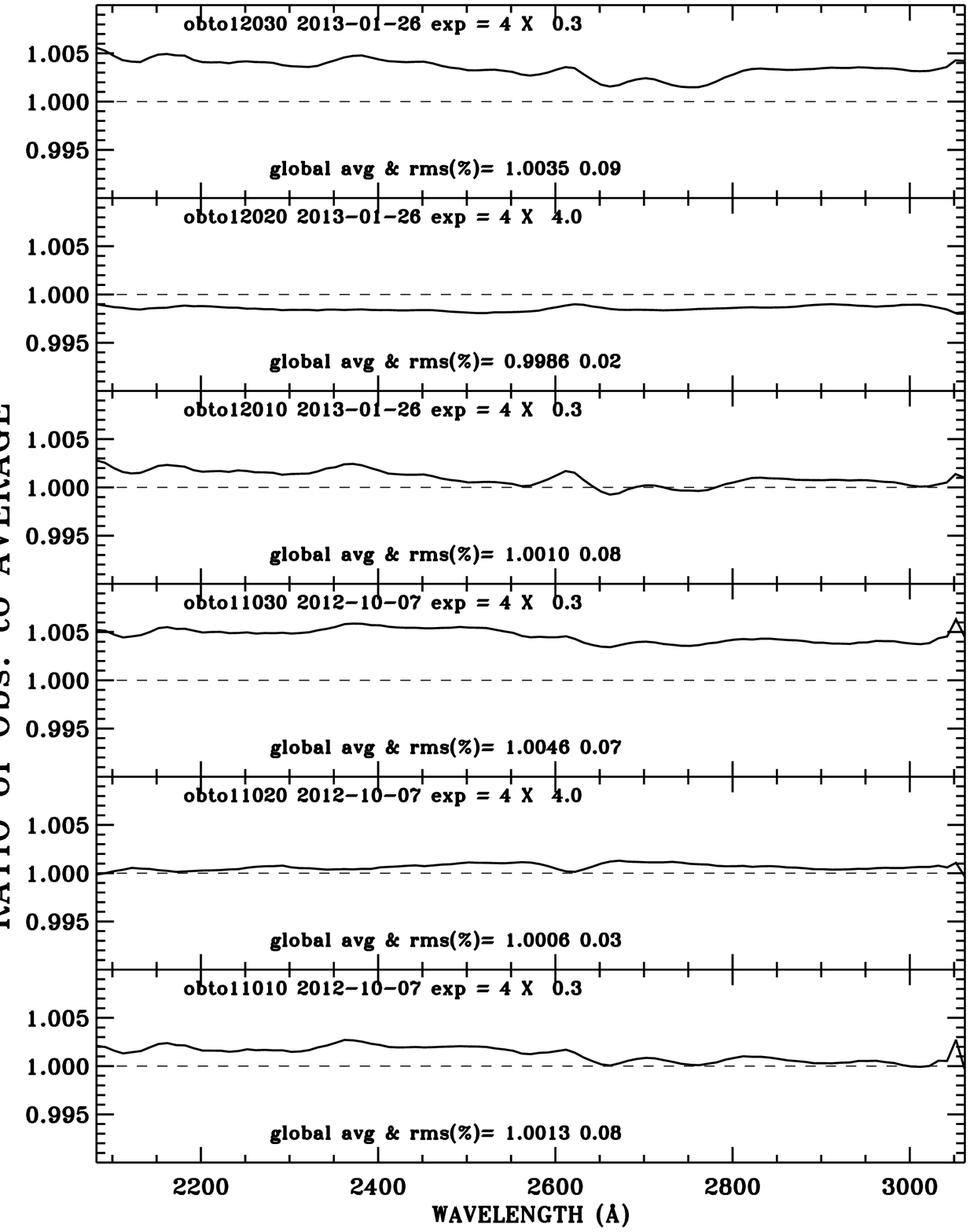}
\caption{\baselineskip=14pt
Ratios of each of the six observations of Sirius in the G230LB mode to the
average of these six spectra. In addition to the exposure time of each of the
CR-split=4 observations, the global average and rms of the residuals are written
on the plots. The average spectrum is dominated by the two heavily saturated
observations of 16s. Each independent observation repeats to $\pm\sim$0.1\%,
except for the last observation of each visit, which is high by $\sim$0.4\%.
The exposure time of the short observations is a small fraction of the total, so
that this small non-repeatability has a $<$0.1\% effect on the final average.
\label{modcf}}
\end{figure}

Table~1 is the journal of the Sirius observations, while Figure~\ref{modcf}
demonstrates both the repeatability of G230LB observations and the linearity
beyond saturation. The individual sub-exposure times from Table~1 are either
0.3s or 4s. Figure~\ref{modcf} shows the  ratio of the six G230LB observations
to their average. The two 16s exposures with four sub-exposures of 4s repeat to
0.2\% and dominate the average spectrum. The 0.3s exposures average 0.30\%
higher than the 4s exposures, in agreement with BG, who also found 0.3009s for
the nominal 0.3s exposure time. However, the scatter of $\sigma=0.17\%$ means
that the 0.30\% exposure time increase has less than a 2$\sigma$ significance;
and 0.3000s is used for the short exposure time. 

%Adopting 0.3009s makes the WD primaries 0.3% brighter, as Vega flux is 
%unchanged. but the Vega G430 & G750L ct/s are less and the WDs are less faint. Sirius flux for
%the 0.3s G430 & G750L exps is unchanged as larger cal fluxes compensates fainter ct/s!!
%Overlap of G230LB & G430L looks good already to ~0.06% & G230LB would be 0.36%
%too bright w/ brighter fluxes per the.3009s exp time change for Vega.

After extracting the spectra from the images, adjusting the flux to a standard
7-pixel-high aperture (Bohlin 1998), and correcting for sensitivity changes with
time using the method of Stys et al. (2004), corrections to the wavelengths are
made for sub-pixel wavelength errors that are obvious in the high S/N saturated
spectra. These shifts range up to 0.79~pixel and are found by cross-correlation
of the absorption lines with a model flux distribution.

\subsection{The new Rauch Flux Calibration}  %2.1

The STIS absolute flux calibration is based on model atmosphere SEDs for the
three primary WDs G191B2B, GD153, and GD71. Gianninas et al. (2011,
G11) fit new Balmer line observations of these WDs with updated NLTE,
pure-hydrogen models that include improved theoretical calculations of the
Balmer lines (Tremblay \& Bergeron 2009). G11 found $T_\mathrm{eff}$ and $\log
g$ of 60920~K and 7.55~$cm~s^{-2}$ for G191B2B, 40320~K and 7.93~$cm~s^{-2}$ for
GD153, and 33590~K and 7.93~$cm~s^{-2}$ for GD71. For G191B2B, RWBK computed
line-blanketed NLTE models and reported a best fit to the absorption lines in
STIS and FUSE high dispersion spectra of $T_\mathrm{eff}=60000 \pm 2000K$ and
$\log g=7.60$. However, a $T_\mathrm{eff}=59000~K$ model is within the
uncertainty and is more consistent with the STIS relative UV flux among the
three stars. In addition, RWBK found  $N(HI)=2.2\times10^{18}~atoms~cm^{-2}$
from the Lyman lines, which corresponds to E(B-V)=0.0005 according to the
galactic average $N(HI)/E(B-V)=4.8\times10^{21}~atoms~cm^{-2}~mag^{-1}$ of
Bohlin et al. (1978).

New models for the three fundamental primary standards GD71, GD153, and G191B2B
(Bohlin 2003) are calculated with the T\"{u}bingen NLTE Model-Atmosphere Package
(Werner et al. 2003, Rauch \& Deetjen 2003), which includes metal line
blanketing for G191B2B but only pure hydrogen for GD153 and GD71 at the G11
$T_\mathrm{eff}$ and $\log g$. The model parameters for the three primary WDs
appear in Table 2. Their SEDs are available via
CALSPEC\footnote{http://www.stsci.edu/hst/observatory/crds/calspec.html} and
also from the
registered Virtual Observatory service TheoSSA\footnote{Theoretical Stellar
Spectra Access, \url{http://dc.g-vo.org/theossa}} that was created in the
framework of the  GAVO\footnote{German Astrophysical Virtual Observatory, 
\url{http://www.g-vo.org}} project. After reddening the G191B2B model by
E(B-V)=0.0005 and normalizing relative to the 5556~\AA\ flux of for Vega, these
three SEDs define the STIS absolute flux calibration and all observed STIS
fluxes.

Previous to November 2013, pure hydrogen models calculated with the Hubeny NLTE
code defined the SEDs of all three stars. The switch to Rauch models results in
a wavelength dependent shift of the HST flux scale by  $<\sim$1\% in the
STIS wavelength range. At 8~\micron, the worst discrepancy with the 
Spitzer/IRAC fluxes (Bohlin et al. 2011) had been a 4$\sigma$ difference of 12\%
for G191B2B. The 4\% lower flux at 8~\micron\ for the new G191B2B SED reduces
the discrepancy to 8\% with less than a 3$\sigma$ significance.

The final Sirius fluxes adjusted to the new calibration can be found in the
CALSPEC database with the file name sirius\_stis\_001.fits. Included in the file
are the estimated systematic uncertainty of 1\% and the statistical
uncertainties, which determine a S/N per pixel that ranges up to 15000 at
4000-4100~\AA, where  $2.4\times10^{8}$ electrons per pixel are extracted from
the CCD spectral images.

\section{COMPARISON OF THE STIS FLUX FOR SIRIUS WITH MODELS AND OTHER RESULTS} %3

\subsection{Kurucz, CWW, and EPK Models}

\begin{figure}%2
\centering
\includegraphics[height=7in,trim=50 30 0 40]{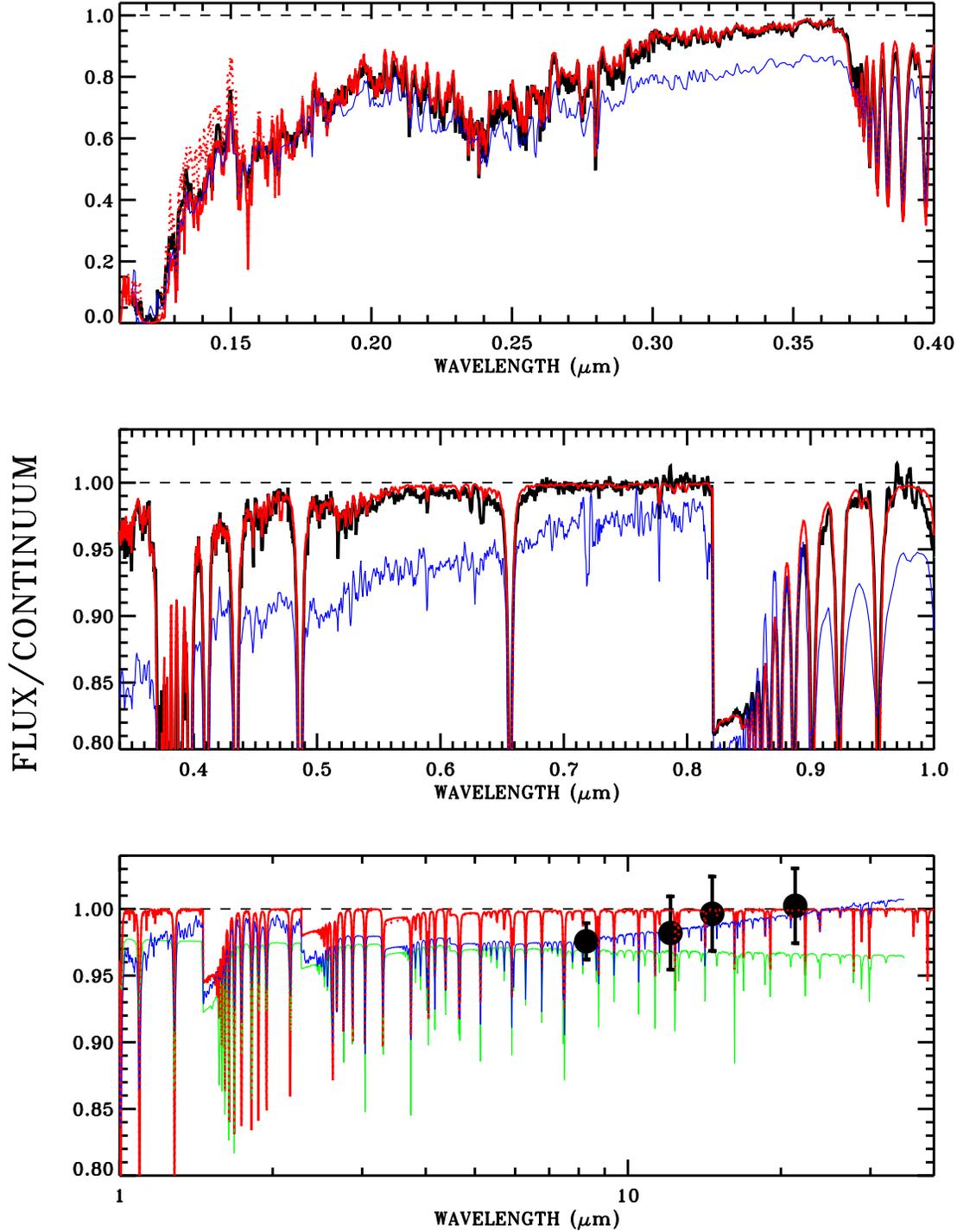}
\caption{
Comparison of STIS (black) to the Kurucz model (red: solid--2013 update, 
dots--original) with $T_\mathrm{eff}=9850~K$, $\log g=4.30$, and $[M/H]=+0.4$.
The original and 2013 Kurucz models coincide, except where the dots are
discernable. The  blue curve is from EPK, while green represents the CWW
template for Sirius. The STIS data cover the wavelength range below 1~$\mu$m and
are supplemented by IUE  data below 1675~\AA. The short wavelength limit of the
CWW SED is 1~$\mu$m. The black filled circles with $2\sigma$ error bars are the
MSX values of PPEM. \label{siriuscont}}   \end{figure}

Figure~\ref{siriuscont} compares the STIS flux for Sirius to other results,
including an $R=500$ resolution, original Kurucz
model\footnote{http://kurucz.harvard.edu/stars/SIRIUS/} and an update (Kurucz
private comm. 2013). All of the illustrated SEDs have been divided by the same
theoretical smooth continuum, as normalized to the STIS flux at 6800--7700~\AA,
in order to display the spectral features on an expanded scale. Ratio plots of
STIS/model have large and distracting, spurious dips and spikes near strong
absorption line features because of small mismatches in resolution and tiny
wavelength errors. The ratios of flux/continuum in Figure~\ref{siriuscont}
display the nature of the absorption lines, and the irrelevant small mismatches
at line centers are often off-scale and can be easily ignored. Longward of
1~\micron, the 2013 Kurucz model, normalized to the STIS flux in the
6800--7700~\AA\ range, is chosen for the composite Sirius SED. Many of the weak
STIS features which are $<$1\% correspond to spectral features in the Kurucz
model. To define the final composite Sirius SED below 1675~\AA, an IUE spectrum
is used after normalization to the STIS flux by multiplying by 1.28. The highest
speed IUE trail data used for bright stars often missed the exact slit center,
making the absolute flux low; but see Section 3.4 for verification of the 1.28
factor. The blue curve in Figure~\ref{siriuscont} is from EPK, while the Cohen
et al. (1992a) reference SED is green. The Cohen SED for Sirius is part of the
CWW standard star network (Cohen et al. 1992b, Cohen 2007).

The EPK curve (blue) is more than 10\% low at places in the top two panels of
Figure~\ref{siriuscont} but differs from the Kurucz extrapolation (red) by only
a few percent in the bottom panel. The EPK fluxes are the zero-point SED from
their table 4, as scaled up by their -1.368 mag for Sirius. Below 0.9~\micron,
the EPK SED is based on the vintage absolute flux for 109  Vir from T\"{u}g et
al. (1977), as scaled to zero magnitude. Between 0.9 and 9~\micron, EPK warp
NICMOS and ISO measured flux distributions to match various photometry. From
9.4--35~\micron, the upward trend of the blue  curve in Figure~\ref{siriuscont}
reflects the EPK smooth adjustment of the CWW model to closely track the MSX
absolute fluxes of Price et al. (2004, PPEM).

% What Teff needed to fit upward swoop of Price matching? // BB calc: planck.pro
% EPH 22/8 mic slope is ~half way btwn BB & Kz model. See epkvsplanck.pro

The normalized Kurucz model (red) agrees with the observed STIS flux (black) to
better than $\sim$1\% over most of the wavelength range in the middle panel of
Figure~\ref{siriuscont}. The original 1993 specially tailored Kurucz 
model\footnote{http://kurucz.harvard.edu/stars/SIRIUS/} agrees with his 2013
update to $<$1\% longward of 1800~\AA; and at shorter wavelengths, the update
fits the observed flux significantly better. This precise agreement of STIS and
model suggests that the modeled flux could represent the true stellar flux with
a similar accuracy of $\sim$1\% longward of 1~\micron.

\subsection{MSX Absolute IR Flux}

Absolute IR flux measurements with direct reference to laboratory standards 
were pioneered from the ground by Selby et al. (1983), Blackwell et al. (1983),
Mountain et al. (1985), and Booth et al. (1989). From space where the atmosphere
does not cause complications, the SPIRIT III instrument on the Midcourse Space
Experiment (MSX) provided pedigreed absolute mid-IR fluxes from emissive
reference spheres (ERS) that were ejected and observed as point sources with
fluxes based on lab data and basic physics. Preliminary MSX results appear in 
Cohen et al. (2001), while PPEM present a definitive final analysis. In their
table 9, PPEM compare their new ERS based absolute fluxes to the ensemble
average of a subset of the CWW network of standards. However, the uncertainties
for each of the 4 MSX bands in that table 9 seem to be the internal statistical
precision to which the MSX flux scale can be related to the CWW average scale.
For the final MSX external uncertainties of the absolute flux measures, the
1.4\% from the PPEM abstract is most appropriate. Larger uncertainties of 3--4\%
are quoted for the calculated ERS fluxes; but those values are reduced by an
analysis of the band-to-band flux ratios, which constrain the sphere temperature.
A minimum error bar arises from the two uncertainties of 2\% attributed to
$\pm$1\% in the sphere radii and $\pm$1\% in the distance to the ejected
spheres. If each of the 5 ERS ejection events are independent, eg. the 1\%
radius error is random and not the same for all 5 released spheres, then the
error-in-the-mean floor for each band is $\sqrt{(4 + 4)/5)} = 1.3\%$, which is
consistent with the adopted 1.4\% in Table 3. 

In addition to the systematic corrections to the global CWW flux scale, the CWW
flux for Sirius itself was found to be offset by 1\% low compared to the
ensemble set of CWW program stars. Table 3 summarizes these results and includes
the correction factors and final MSX in-band, effective fluxes (irradiance) for
Sirius, where the negative corrections mean that the CWW flux is low and must be
increased by the tabulated factors to agree with the MSX results. The fluxes in
column 5 of Table 3 are the integral fluxes over the bandpass from table 2 of
PPEM divided by the table 1 bandwidth of Cohen et al. (2001) times the
correction in column 4 of Table 3. Multiplying this total CWW correction factor
by the column 7 ratio of the green CWW to the red curve adopted for the IR
extension of the HST/STIS fluxes in Figure~\ref{siriuscont} produces the ratio
of MSX to the adopted IR flux. These ratios of MSX to HST/STIS are tabulated in
the final column (8) of Table 3 and are plotted with their 2$\sigma$ error bars
of 2.8\% as the filled circles at the MSX isophotal wavelengths in
Figure~\ref{siriuscont}.

To normalize the Kurucz model to the MSX lab based absolute flux calibration,
the illustrated normalization that is based on the 5556~\AA\ Vega flux of
Megessier (1995) would be multiplied by the average of the four factors in the
final column of Table 3. However to properly weight the visible and mid-IR
absolute fluxes, a weighted average correction of $0.9945\pm0.5$\% includes the
5556~\AA\ normalization at unit value $\pm0.7$\% with a $1/0.7^2$ weight along with
the four IR corrections with their $1/1.4^2$ weights. Equivalently, to establish
a new HST flux scale that is based on Vega at 5556~\AA, the Megessier value
should be multiplied by 0.9945. A revised, MSX weighted value of
$3.46\times10^{-9}\times0.9945 = 3.44\times10^{-9}\pm0.5$\% is adopted.
(Coincidentally, this new value is the same as recommended by Hayes (1985).) The
most significant deviation among the five absolute flux measures is for the MSX
A band, where the 0.976 offset from Table 3 is improved by the 0.9945 factor to
0.982, i.e. a deviation from unity by 1.8/1.4 = 1.3$\sigma$. The revised
absolute normalization is within the uncertainty expectations of all five
absolute flux measures from the visible to the mid-IR.

\subsection{Linnell and M\'esz\'aros Models}

\begin{figure}%3
\centering
\includegraphics[height=7in,trim=50 0 0 40]{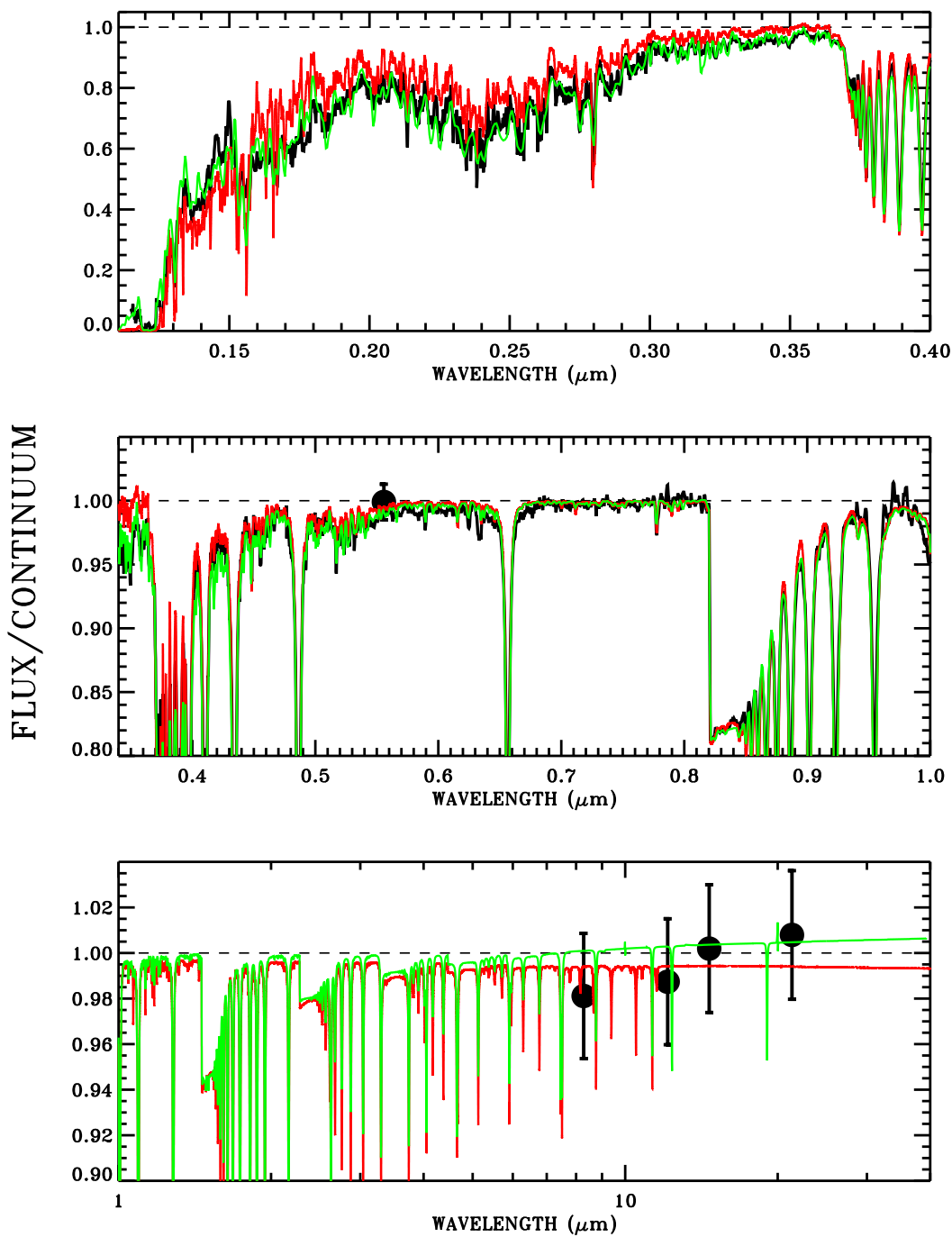}
\caption{
Comparison of STIS (black) to the best Sirius models by Linnell
(green) and M\'esz\'aros (red), as in Figure~\ref{siriuscont}. The black filled
circles are the MSX values of PPEM but are now shown after the 0.9945 correction
to the HST/STIS flux scale. The black circle at 0.556~\micron\ is the
$3.46\times10^{-9}$~erg cm$^{-2}$ s$^{-1}$ \AA$^{-1}$ value of Megessier (1995),
which is now 0.55\% above the STIS flux. Notice the expanded scale in the bottom
panel, where both models are within 1\% of the reference Kurucz model (red in
Figure~\ref{siriuscont}).  \label{lnmzcont}}   \end{figure}

While the Sirius flux is difficult to model in detail in the far-UV, where the
line-blanketing is severe, other modeling efforts reproduce the visible--IR
shape. Figure~\ref{lnmzcont} compares the best efforts of Linnell et al. (2013
and private comm.) and of M\'esz\'aros et al. (2012 and private comm.) to the
STIS fluxes by dividing by the same Kurucz continuum level as in
Figure~\ref{siriuscont}. The Linnell model shown in Figure~\ref{lnmzcont} is
slightly improved over the on-line SED. The small differences between the 
Kurucz and M\'esz\'aros models are probably due to different solar abundances
and different updates to the synthesis code. M\'esz\'aros uses the Asplund et
al. (2005) abundances and the F. Castelli version of the synthesis code, while
Linnell uses the I. Hubeny line list and Synspec. The modeling of the Sirius IR
SED is robust with all three independent models agreeing to better than
$\pm$1\%,  which increases confidence that the adopted Sirius IR flux is
accurate to 1\%.

\subsection{SOLSTICE Absolute UV Flux}

\begin{figure}%4
\centering
\includegraphics[height=7.5in,trim=50 0 0 0]{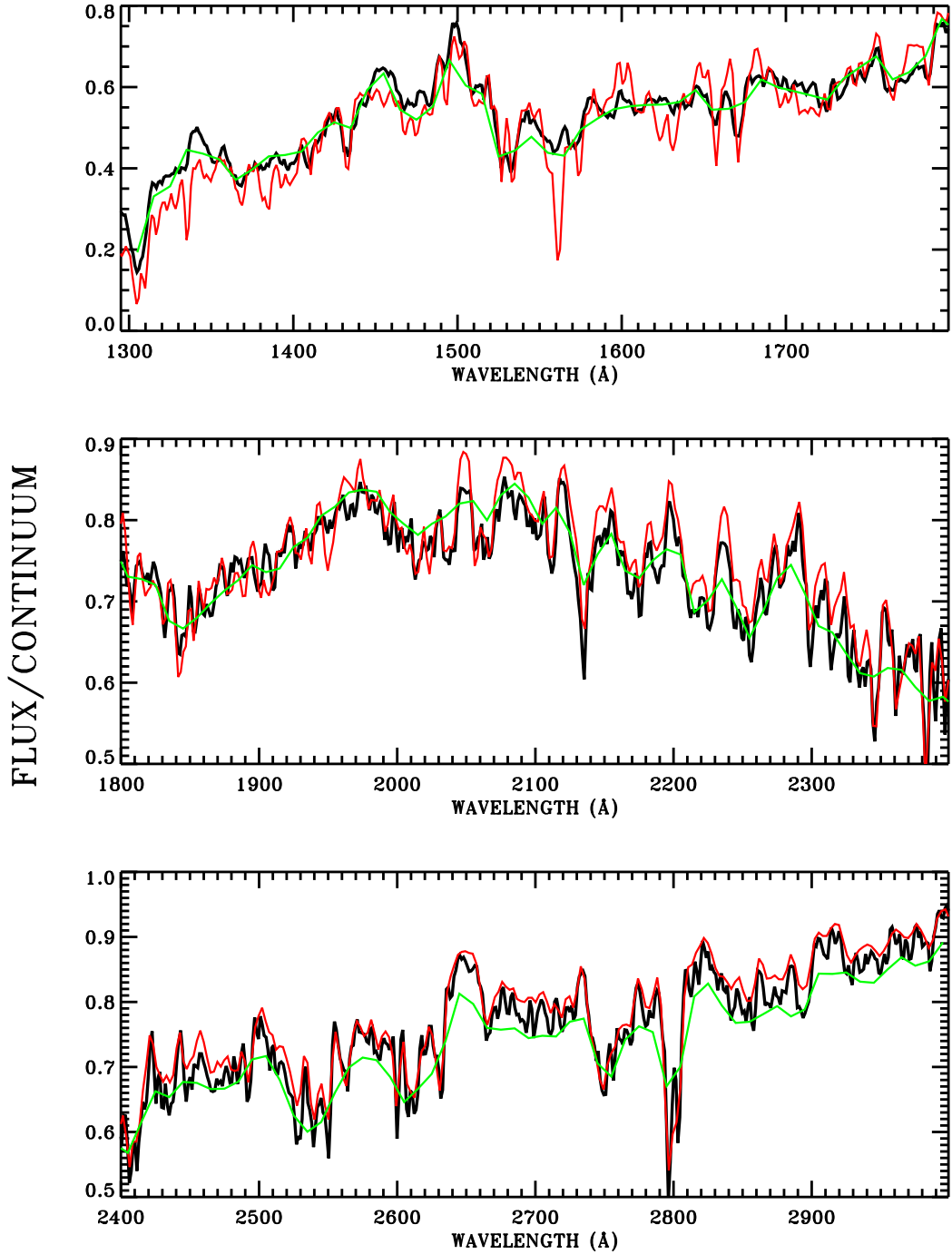}
\caption{
Comparison of STIS+IUE (black) and the Kurucz Sirius model (red) to the
independent lab based SOLSTICE absolute fluxes (green), as in
Figure~\ref{siriuscont}. The two observational results agree to $\sim$3\%, 
while the model tracks both observations with remarkable fidelity, considering
the computational difficulty of accounting for the heavy line-blanketing at
these UV wavelengths. \label{solsticecont}}   \end{figure}

Figure~\ref{solsticecont} shows that the absolute UV flux of Sirius that is
defined by IUE and STIS is confirmed by the independent lab based calibration of
Snow et al. (2013) for the SOLSTICE fluxes\footnote{http://bdap.ipsl.fr/fondue/}
(green) over their 1300--3000~\AA\ range. These two measured absolute flux
levels agree to $\sim$3\%, as expected from the SOLSTICE/STIS comparison for
three stars in Bohlin \& Gordon (2014). The structure of the spectral features
in the STIS data (black) at R$\sim$500 often matches the Kurucz R=500 resolution
model (red) longward of 1675~\AA. The R$\sim$250 IUE data (black) below
1675~\AA\ and the R$\sim$100 SOLSTICE SEDs also nicely track the flux level of
the model, but the spectral structure is smoother than the R=500  model. In
general, the observed spectral structure is remarkably well reproduced by the
theoretical modeling, even with the heavy UV line-blanketing.

\section{RESULTS FOR VEGA}

\subsection{Dust Rings}

\begin{figure}%5
\centering
\includegraphics[height=6in,trim=50 0 0 0]{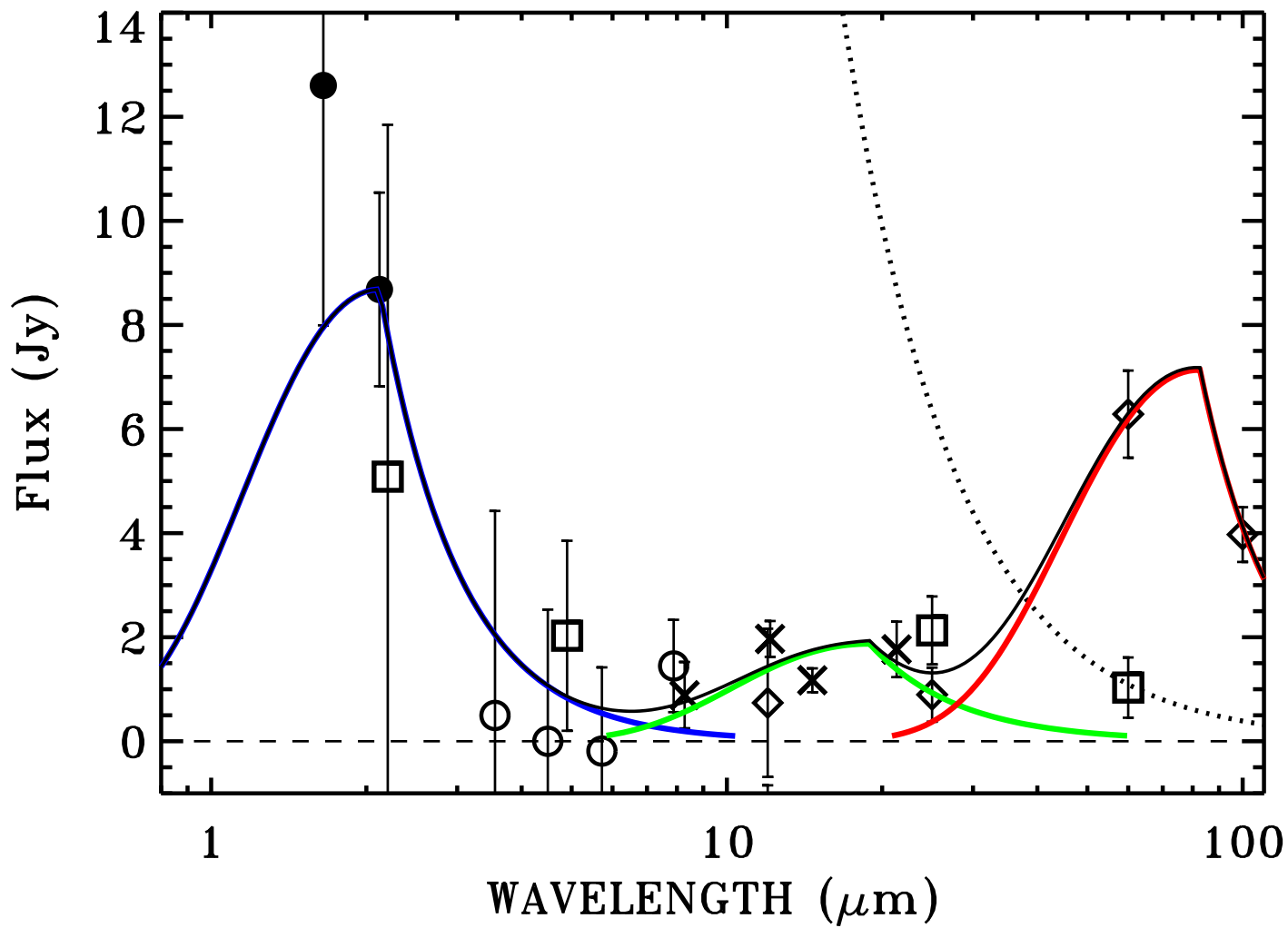}
\caption{
Emission from the Vega dust rings (data points and solid black line fit), which
is the total flux minus the CALSPEC photospheric SED. The dotted line is the
CALSPEC SED, i.e. the Kurucz photospheric model normalized to STIS at
6800--7700~\AA. The data points with their 1$\sigma$ error bars are from various
IR space missions: squares (DIRBE), open circles (IRAC), $\times$ (MSX),
diamonds (IRAS); and the filled circles are the ground-based interferometric
results. The individual dust ring components are for 2400~K (blue), 270~K
(green), and 62~K (red), while the continuous black line is the sum of these
three components. \label{vegadust}}   \end{figure}

Once Sirius is established as a primary flux standard, the flux of any other
star is determined by just the brightness relative to Sirius. These signal
ratios are free of errors in instrumental flux calibrations. For example, the
ratio of Vega to Sirius is available for a few IR missions; and the total flux
of Vega on the HST absolute flux scale is this ratio times the well-modeled flux
of the primary IR standard, Sirius. Vega is a pole-on rapid rotator (Peterson
et~al. 2006) requiring a model with temperature zones in the 7900--10150~K range
(Aufdenberg et~al. 2006). However longward of 4000~\AA, a single temperature
Kurucz model at 9400~K fits the STIS SED to 1\%, as discussed by Bohlin (2007).
After subtracting this CALSPEC Kurucz model for the Vega photospheric flux
($T_\mathrm{eff}=9400~K$, $\log g=3.90$, $[M/H]=-0.5$, Bohlin 2007), the
remainder is the emission from the dust rings surrounding Vega, as shown in
Figure~\ref{vegadust}. The photospheric Kurucz model that fits the STIS flux for
Vega is the dotted line, which decreases steeply to less than the total dust
emission beyond $\sim$40~\micron.

The measured ratios of Vega/Sirius include the four broadband MSX points of PPEM
($\times$), i.e. the Vega flux from their table 3 divided by the Sirius flux
from their table 2. The open circles in Figure~\ref{vegadust} are from the IRAC
ratios (Marengo et al. 2009) with their 1\% error bars. The DIRBE  data are from
the electronic version of table 1 in Smith et al. (2004), while Neugebauer et
al. (1984) review the  IRAS
data\footnote{http://irsa.ipac.caltech.edu/cgi-bin/Gator/nph-dd}. Three of the
six DIRBE values (squares) in Figure~\ref{vegadust} are low at 3.5, 12, and
60~\micron, even after correcting Sirius by 0.014 mag for other stars in the 
42~arcmin DIRBE beam (Su et al. 2013, hereafter Su). The 3.5 and 12~\micron\
points are offscale in Figure~\ref{vegadust}, while the three DIRBE measures at
2.2, 4.9, and 25~\micron\ nicely track the other experimental results. The 
error bars on the DIRBE points are the rms scatter divided by the square root of
the number of observations (N); however, there is another quoted DIRBE
uncertainty $\langle err\rangle$ due to background fluctuations. What is not
clear is whether or not $\langle err\rangle$ should be reduced by $\sqrt{N}$ to
get total DIRBE uncertainties. The illustrated error bars for DIRBE, IRAC, MSX,
and IRAS all include a 1\% uncertainty for the Sirius flux combined in
quadrature with the instrumental rms scatter.

Using interferometric techniques in the K band at 2.1~\micron, Absil et al.
(2013) measure a 1.26$\pm0.27$\% dust contribution with respect to the Vega
photosphere. Similarly, Defr{\`e}re et al. (2011) find a 1.23$\pm 0.45$\%
contribution from small grains at the H band. These interferometric results are
the filled black circles in Figure~\ref{vegadust} and are independent of the
Sirius SED.

Su models the dust with three rings at different temperatures: An inner hot ring
of radius $\sim1.5\arcsec$ at 2400~K, a warm 170~K ring at 2--3\arcsec, and a
cold 50~K ring peaked at 11--14\arcsec. For the data analyzed here, a 270~K
blackbody fits the warm component better than the 170~K of Su, while 62~K is used
for the cold dust. Planck blackbody curves for the three temperatures are
normalized to the observed fluxes in Figure~\ref{vegadust}. Following Su, the
three blackbody curves are truncated longward of the peak with a steep
$\lambda^{-x}$ decline, where x=2.8, 2.5, and 2.9, instead  of the x=2 for the
Rayleigh--Jeans slope. The total emission from all three blackbody curves
(black curve) falls within $\sim1\sigma$ of the observational data, except
for a $\sim2\sigma$ deviation for the MSX 14.65~\micron\ point and for the two
offscale and the 60~\micron\ DIRBE points.

The normalization of the three dust components is summarized in Table 4
and includes the ratio of the peak dust emission to photosphere in the final
column. The sum of dust emission plus photosphere could be used as a standard
star SED in the IR with a $\sim$1\% photospheric uncertainty plus the tabulated
uncertainty in the dust contribution. However, this total IR Vega flux is 
dependent on the observational aperture size, which compromises its practical
utility as a standard star longward of 1~\micron.

The Kurucz special photospheric models that do not include the dust
contributions are in the CALSPEC database as
alpha\_lyr\_mod\_001.fits ($T_\mathrm{eff}=9400~K$, $\log g=3.90$,
$[M/H]=-0.5$) and sirius\_mod\_001.fits ($T_\mathrm{eff}=9850~K$, $\log g=4.30$,
$[M/H]=+0.4$). 

\subsection{No \emph{Hipparcos} Evidence for Variability of Vega}

\begin{figure}%6
\centering
\includegraphics[height=5.0in,trim=50 0 0 0]{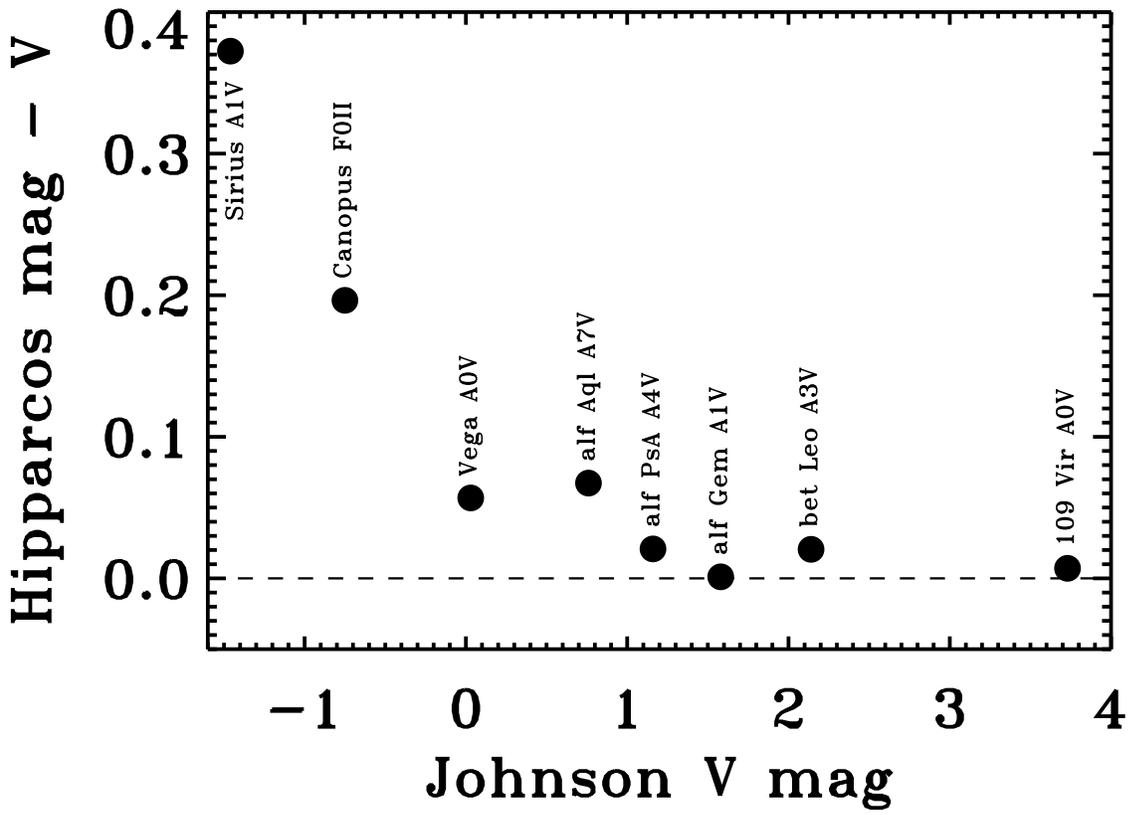}
\caption{Non-linearity of \emph{Hipparcos} $H_p$ magnitudes for the brightest 
A and F stars.
\label{hiplin}}  
\end{figure}

EPK reported that the \emph{Hipparcos} photometry $H_p$ (van Leeuwen et al.
1997) shows that Vega is a variable star. Those data are reviewed here with the
conclusion that this apparent variability is most likely caused by saturation of
the pulse counting electronics in the image dissector photomultiplier tube. For
the brightest stars, Figure~\ref{hiplin} shows increasing differences between 
$H_p$\footnote{http://vizier.u-strasbg.fr/viz-bin/VizieR-3?-source=I/239/hip\_main}
and the V mag (Johnson et al. 1966) for A and F stars. For example, $H_p$ for
Sirius is $\sim$0.37 mag too faint, while for stars fainter than V=1, the two
magnitude measures are equal within $\sim$0.02. For Vega, the \emph{Hipparcos}
0.087 mag is 0.06 fainter than the standard Johnson value of V=0.03.

\begin{figure}%7
\centering
\includegraphics[height=5.0in]{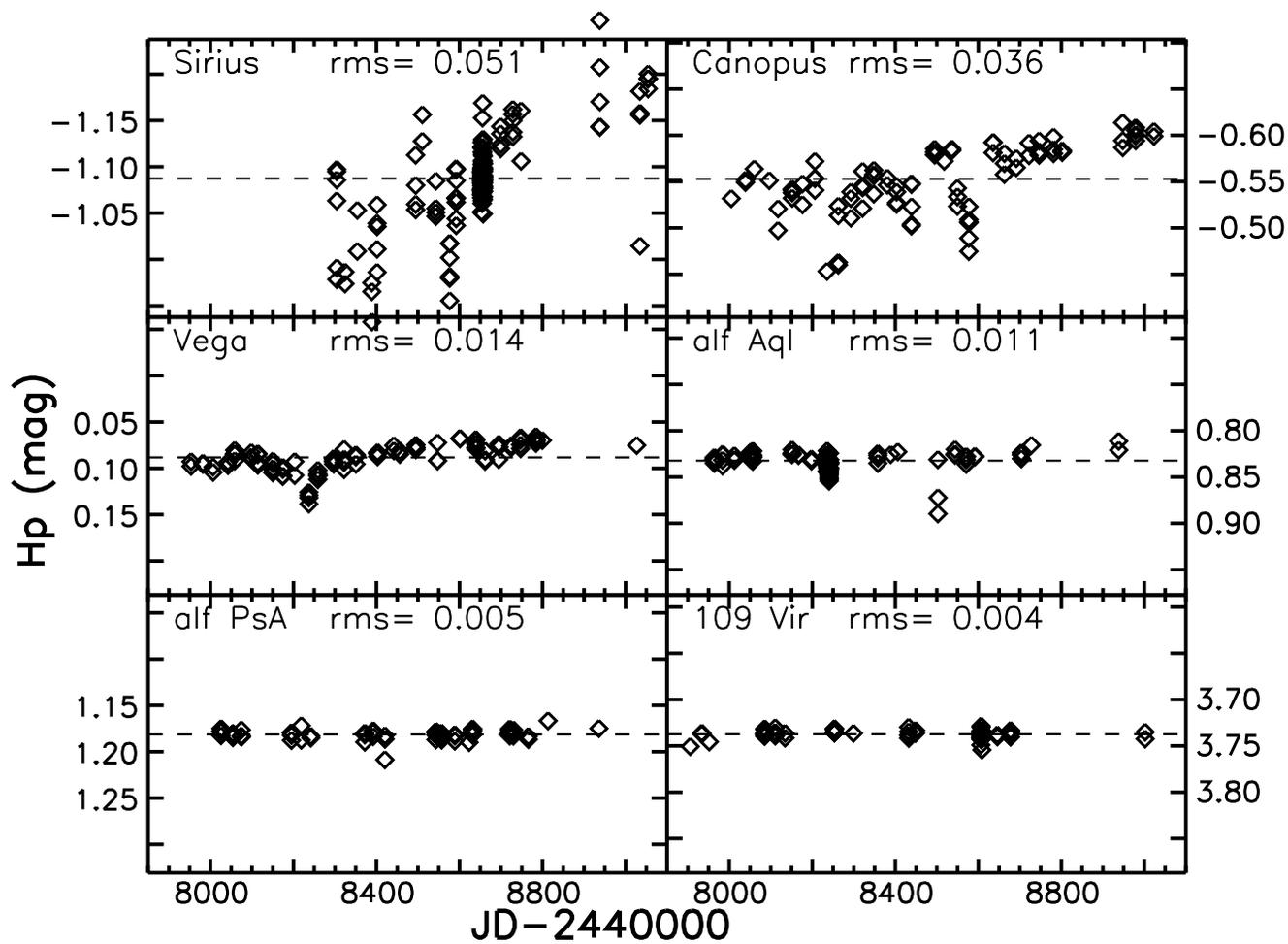}
\caption{Apparent variability of \emph{Hipparcos} $H_p$ magnitudes for bright
stars from a Johnson V mag of -1.46 for Sirius in the uppper left panel to
V=3.73 for 109 Vir at the lower right. The brightest stars with $V<1$ in the
top four panels are affected by instabilities at these highest count rates where
pulse pileup is  significant.
\label{hipvari}}  
\end{figure}

Figure~\ref{hipvari} shows that the exact effect of the pulse pileup is
sensitive to environmental conditions, eg. temperature, as the electronics
attempt to discriminate between coincident pulses. Thus, the recorded count rate
fluctuates the most for the brightest star, Sirius, where the rms scatter of the
160 separate measures is 0.051 mag. For Vega after removing a spurious 2.29
value, the rms of the remaining 102 values is 0.014 mag, i.e. considerably above
the typical 0.004--0.005 mag rms of fainter stars, where the $H_p$ photometry
becomes linear.

EPK state that while the discrepant magnitude for Vega "... might be
attributed to saturation effects, other, brighter stars do not show such
discrepancies between V and $H_p$. Hence, we conclude that the faintness of Vega
cannot be attributed to non-linearity in \emph{Hipparcos}." This work
demonstrates just the opposite: Brighter stars $are$ discrepant and the faint
$H_p$=0.087 for Vega  $is$ caused by non-linearity. Furthermore, the apparent
variability and gradual apparent brightening of the four brightest stars in
Figure~\ref{hipvari} is probably caused by slight variations in the exact level
of pulse pileup in the detector. The \emph{Hipparcos} photometry does not
provide good evidence for the variability of Vega.

\acknowledgements

Constructive comments on preliminary drafts were provided by M. Bessel, S.
Deustua, R. Kurucz, Sz. M\'esz\'aros, T. Rauch, and the referee. Primary support
for this work was provided by NASA through the Space Telescope Science
Institute, which is operated by AURA, Inc., under NASA contract NAS5-26555.

\clearpage
\begin{deluxetable}{llllcc}
% Table1
%\rotate
\tablewidth{0pt}
\tablecaption{Journal of STIS Observations in the 52X2 Arcsec Slit}
\tablehead{
\colhead{Root} &\colhead{Mode}
&\colhead{Date}
&\colhead{Time} &\colhead{Exptime~(s)\tablenotemark{a}}&\colhead{Repeats}
}
\startdata
obto11010  &G230LB  &12-10-07 &21:03:13    &  1.2  &4\\
obto11020  &G230LB  &12-10-07 &21:06:13    & \llap{1}6.0  &4\\
obto11030  &G230LB  &12-10-07 &21:12:22    &  1.2  &4\\
obto11040  &G430L   &12-10-07 &21:23:12    &  0.9  &3\\
obto11050  &G430L   &12-10-07 &21:25:27    &  0.9  &3\\
obto11060  &G750L   &12-10-07 &21:35:32    &  2.1  &7\\
obto12010  &G230LB  &13-01-26 &12:10:59    &  1.2  &4\\
obto12020  &G230LB  &13-01-26 &12:13:59    & \llap{1}6.0  &4\\
obto12030  &G230LB  &13-01-26 &12:20:08    &  1.2  &4\\
obto12040  &G430L   &13-01-26 &12:30:58    &  0.9  &3\\
obto12050  &G430L   &13-01-26 &12:33:13    &  0.9  &3\\
obto12060  &G750L   &13-01-26 &12:43:18    &  2.1  &7\\
\enddata
\tablenotetext{a}{Total exposure time. Individual sub-exposure 
integration times are
this total divided by the number of repeats in the final column.}
\end{deluxetable}

%2
\begin{deluxetable}{cccccc}
%\tabletypesize{\scriptsize}
\tablewidth{0pt}
\tablecolumns{6}
\tablecaption{The Primary WD Stars}
\tablehead{
\colhead{Star} &\colhead{V\tablenotemark{a}} &\colhead{Sp. T.} 
&\colhead{$T_\mathrm{eff}$} &\colhead{$\log g$}
&\colhead{Unc.$T_\mathrm{eff}$}}
\startdata
G191B2B &11.781 &DA.8  &59000 &7.60 &2000 \\
GD153   &13.346 &DA1.2 &40320 &7.93 &~626 \\
GD71    &13.032 &DA1.5 &33590 &7.93 &~483 \\
\enddata
\tablenotetext{a}{G191B2B--Landolt and Uomoto (2007), GD153--Landolt private
comm, GD71--Landolt (1992)}

\end{deluxetable}

%3
\begin{deluxetable}{cccccccc}
%\tabletypesize{\scriptsize}
\tablewidth{0pt}
\tablecolumns{8}
\tablecaption{Sirius Flux Measured by MSX}
\tablehead{

&\multicolumn{3}{c}{----------- Corrections -----------} 
&&&\multicolumn{2}{c}{------ Corrections -----}\\

\colhead{Band} &\colhead{Avg CWW} &\colhead{Sirius} 
&\colhead{Total CWW} &\colhead{Flux\tablenotemark{a}} &\colhead{Uncert}
&\colhead{CWW/HST} &\colhead{HST}\\

\colhead{(\micron)~~~~} &\colhead{(\%)} &\colhead{(\%)} &\colhead{factor}
&&\colhead{(\%)}
&\colhead{factor}&\colhead{factor}}

%eg: 1000{W-->erg}*2.833e-15{price-Table2}/3.362{Cohen-Table1}/.994
% see cwwvshst.pro

\startdata
~8.28 (A) &+0.4 &-1.0 &1.006 &8.477E-13 &1.4 &0.970 &0.976 \\
12.13 (C) &-0.4 &-1.0 &1.014 &1.883E-13 &1.4 &0.968 &0.982 \\
14.65 (D) &-1.9 &-1.0 &1.030 &9.031E-14 &1.4 &0.968 &0.996 \\
21.34 (E) &-2.5 &-1.0 &1.036 &2.025E-14 &1.4 &0.967 &1.002 \\
\enddata
\tablenotetext{a}{erg~cm$^{-2}$~s$^{-1}$~\AA$^{-1}$}
\end{deluxetable}

%4
\begin{deluxetable}{ccccc}
%\tabletypesize{\scriptsize}
\tablewidth{0pt}
\tablecolumns{5}
\tablecaption{Parameters of Vega Dust Model}
\tablehead{

\colhead{$T_\mathrm{eff}$} &\colhead{$\lambda_{peak}$} &\colhead{Peak Flux} 
&\colhead{Uncert} &\colhead{Peak/Photos}\\

\colhead{(K)} &\colhead{(\micron)} &\colhead{(Jy)} &\colhead{(Jy)} &}

\startdata
2400 &2.1 &8.7 &1.9 &0.013 \\
270  &19  &1.9 &0.5 &0.17 \\
62   &83  &7.1  &1.0 &12.5 \\
\enddata
\end{deluxetable}

\end{document}